\documentstyle{article}
\begin{document}
\large \centerline { SYMMETRIES OF TODA EQUATIONS } \vskip 2 cm
\centerline { Pantelis A. Damianou } \centerline { Department of
Mathematics and Statistics } \centerline  {The University of
Cyprus} \centerline { P. O. Box 537, Nicosia, Cyprus } \vskip 5 cm
\centerline { \bf {ABSTRACT }}
\bigskip
{\it We find a sequence consisting of time dependent evolution vector fields
whose time independent part corresponds to the master symmetries for the Toda
equations. Each master symmetry decomposes as a sum consisting of a group
symmetry and a Hamiltonian vector field. Taking Lie derivatives in the direction
of these vector fields produces an infinite sequence of recursion operators.}
\vfill
\eject
{\bf 1. Introduction: } A symmetry group of a system of differential equations
 is  a Lie group acting on the space of independent and dependent
variables in such a way that solutions are mapped into other solutions.
Knowing the symmetry  group allows one to determine some special types of solutions
 invariant under a subgroup of the full symmetry group, and in some cases
  one can  solve the equations completely.
 The symmetry approach to solving differential equations can
be found, for example, in the books of Olver \cite{olver2}, Bluman
and Cole \cite{cole}, Bluman
 and Kumei \cite{kumei}, and  Ovsiannikov \cite{ovsia}.
One method of finding symmetry groups is the use of recursion
operators, an idea introduced by Olver \cite{olver1}.
  The existence of a recursion operator
provides a mechanism  for generating infinite
hierarchies of symmetries. Most of the well known integrable equations, including
the KdV, do
have a recursion operator. Even some non conservative systems have  recursion operators.
The Toda Lattice is one example where a recursion operator is not known.
 In  \cite{damianou2}  we used master symmetries to generate nonlinear
Poisson brackets for the Toda Lattice.  In essence, it is an example of a system
 which is not only bihamiltonian but it can actually be given $N$ different
 Hamiltonian formulations with $N$ as large as we please.
In most cases, if a system is bihamiltonian, one can find a
recursion operator by inverting one of the Poisson operators.
However in the case of Toda Lattice both operators are
non-invertible and therefore this method fails. Master  symmetries
were first introduced by Fokas and Fuchssteiner in  \cite{fokas1}
in connection with the Benjamin-Ono Equation. Then in W. Oevel and
B. Fuchssteiner \cite{oevel}
 a master symmetry was found for the  Kadomtsev-Petviashvili equation.
 Master symmetries for  equations in $1+1$,  like the KdV, are discussed in  Chen,
  Lee and Lin \cite{chen} and in  Fokas \cite{fokas2}.  General theory of master symmetries is discussed
  in Fuchssteiner \cite{fuch}.   Connection between master symmetries and usual recursion
   operators for equations in $2+1$ is discussed in  \cite{fokas3}.
Some properties of master symmetries ( at least in the Toda case )
are clear: They preserve constants  of motion, Hamiltonian vector
fields and they generate a hierarchy of Poisson brackets. We are
interested in the following problem : Can one find a symmetry
group of the system whose infinitesimal generator is a given
master symmetry? In other words, is a master symmetry a group
symmetry? In the case of Toda equations the answer is negative.
However, in this paper we find a sequence consisting of time
dependent evolution vector fields whose time independent part
corresponds to the master symmetries in \cite{damianou2}. Each
master symmetry $X_n$ can be written in the form $Y_n + t Z_n$
where $Y_n$ is a time dependent symmetry and $Z_n$ is time
independent Hamiltonian symmetry (i.e. a Hamiltonian vector
field). Taking Lie derivatives in the direction of $X_n$ (or
$Y_n$) gives an infinite sequence of recursion operators for the
Toda Lattice.

\vskip 1cm
\bigskip
\noindent
{\bf 2. } In this section we present some background on the Toda lattice. See
 \cite{flaschka} for more details. We also include some of
the results in  \cite{damianou2}  for completeness.

\smallskip

The Toda lattice is a Hamiltonian system with Hamiltonian

 $$ H(q_1, \dots, q_N, \,  p_1, \dots, p_N) = \sum_{i=1}^N { 1 \over 2} p_i^2 +
\sum _{i=1}^{N-1} e^{ q_i-q_{i+1}}  \ .$$

\smallskip
\noindent
This system is completely integrable. One can find a set of functions
$\{  H_1, \dots,  H_N \} $  which are constants of motion for Hamilton's equations.
To  determine the constants of motion, we use Flaschka's transformation
$$
  a_i  = {1 \over 2} e^{ {1 \over 2} (q_i - q_{i+1} ) }  \ \ , \ \ \ \ \ \ \ \ \ \ \ \ \ \
             b_i  = -{ 1 \over 2} p_i  \ .    $$

\smallskip
\noindent
Then

 $$\dot a _i =  a_i (b_{i+1} -b_i )
   \dot b _i =  2( a_i^2 - a_{i-1}^2 ) \ .  $$

\smallskip
\noindent
These equations can be written as a Lax pair  $\dot L = [B, L] $, where $L$ is the
Jacobi matrix

 $$L= \pmatrix { b_1 &  a_1 & 0 & \cdots & \cdots & 0 \cr
                   a_1 & b_2 & a_2 & \cdots &    & \vdots \cr
                   0 & a_2 & b_3 & \ddots &  &  \cr
                   \vdots & & \ddots & \ddots & & \vdots \cr
                   \vdots & & & \ddots & \ddots & a_{N-1} \cr
                   0 & \cdots & & \cdots & a_{N-1} & b_N   \cr } \ , $$

\smallskip
\noindent
and

   $$ B =  \pmatrix { 0 & a_1 & 0 & \cdots & \cdots &  0 \cr
                 -a_1 & 0 & a_2 & \cdots & & \vdots  \cr
                    0  & -a_2 & 0 & \ddots &  & \cr
                    \vdots &  & \ddots & \ddots & \ddots & \vdots \cr
                     \vdots & & &  \ddots & \ddots & a_{N-1} \cr
                     0 & \cdots &\cdots &  & -a_{N-1}  & 0 \cr } \ .$$

\smallskip
\noindent
It follows easily that the eigenvalues of $L$ do not evolve with time.

\smallskip
In  \cite{damianou1} \cite{damianou2}  we constructed a sequence
of vector fields $X_n$, for $n \ge -1$, and an infinite sequence
of contravariant 2-tensors $w_n$, for $n\ge 1$, satisfying :

\smallskip
\noindent
{\it i) } $w_n$ are all Poisson.

\smallskip
\noindent
{\it ii) } The functions $H_n= { 1 \over n} {\rm Tr} \ L^n$ are in involution
 with respect to all of the $w_n$.

 \smallskip
 \noindent
 {\it iii)}  $X_n (H_m) =(n+m) H_{n+m} $.

 \smallskip
 \noindent
{\it iv)} $L_{X_n} w_m =(n-m+2) w_{n+m} $,  modulo an equivalence relation
defined in [].

\smallskip
\noindent
{\it v)} $[X_n, \chi_l] =(l-1) \chi_{l+n} $, where $\chi _l$ is the
Hamiltonian vector field generated by $H_l$ with respect to $w_1$.

\smallskip
\noindent
{\it vi)} $M_n\ {\rm grad }\  H_l =M_{n-1}\  {\rm grad}\  H_{l+1} $,
where $M_n$ is the Poisson matrix  of $w_n$.  If we denote the Hamiltonian
vector field of $H_l$ with respect to the $n$th bracket by $\chi_l^n$, then
these relations are equivalent to $\chi_l^n = \chi_{l+1}^{n-1}$.

\smallskip
We give an outline of the construction of the vector fields $X_n$. We
define $X_{-1} $ to be
$$
 {\rm grad}\  H_1 = {\rm grad } \ {\rm Tr}\ L = \sum_{i=1}^N {\partial \ \over \partial b_i}
 $$
\smallskip
\noindent
and $X_0$ to be the Euler vector field

 $$ \sum _{i=1}^{N-1} a_i {\partial \ \over \partial a_i} +
\sum_{i=1}^N b_i {\partial \ \over \partial b_i} \ .$$

\smallskip
\noindent
We want $X_1$ to satisfy

  $$     X_1 ({\rm Tr}\  L^n) =n {\rm Tr}\  L^{n+1} \ .$$

\smallskip
\noindent
One way to find such a vector field is by considering the equation
\begin{equation}
\dot L = [B, L] +L^2  \ .
\end{equation}
\smallskip
\noindent
Note that the left hand side of this equation is a tridiagonal matrix while the
right hand side is pentadiagonal. We look for $B$ as a tridiagonal matrix
\begin{equation}
B= \pmatrix { \gamma_1 & \beta_1 & 0 & \cdots & \cdots \cr
                \alpha_1 & \gamma_2 & \beta_2 & \cdots & \cdots \cr
                0 & \alpha_2 & \gamma _3 & \beta _3 & \cdots \cr
                \vdots & \vdots & \ddots & \ddots & \ddots \cr}  \ .
\end{equation}

\smallskip
\noindent
 We want to choose the $\alpha_i$, $\beta _i$ and $\gamma_i$ so that the
 right hand side of equation  (9) becomes tridiagonal. One  simple solution
 is \  $\alpha _n = -(n+1) a_n $, \ $\beta _n = (n+1) a_n $,\  $ \gamma_n =0$.
 The vector field $X_1$  is defined by the right hand side of (9) and :
 \begin{equation}
 X_1 = \sum_{n=1}^{N-1} \dot a_n {\partial \ \over \partial a_n} +
   \sum _{n=1}^N \dot b_n {\partial \ \over \partial b_n} \ ,
   \end{equation}
  \smallskip
  \noindent
  where

  \smallskip
  \noindent
  \begin{equation}
  \begin{array}{rcl}
 \dot a_n &= & - na_n b_n + (n+2) a_n b_{n+1} \\

\dot b_n &=& (2n+3) a_n^2 + (1-2n) a_{n-1}^2 + b_n^2  \ .
\end{array}
\end{equation}
\smallskip

  To construct the vector field $X_2$ we consider the equation
  \begin{equation}
 \dot L =[B, L]+L^3 \ .
 \end{equation}
  \smallskip
  \noindent
  The calculations are similar to those for $X_1$. The matrix $B$ is now
  pentadiagonal and the system of equations slightly more complicated. The
  result is a vector field
  \begin{equation}
  X_2 = \sum_{n=1}^{N-1} \dot a _n  {\partial \ \over \partial a_n }+
  \sum _{n=1}^N \dot b _n {\partial \ \over \partial b_n}
  \end{equation}

  \smallskip
  \noindent
  where

  \begin{equation}
  \begin{array}{rcl}
 \dot a_n& =& (2-n)a_{n-1}^2 a_n + (1-n) a_n b_n^2 +a_n b_n b_{n+1} +  \\
     & & (n+1) a_n a_{n+1}^2+ (n+1) a_n b_{n+1}^2 +a_n^3 +\sigma _n a_n (b_{n+1}-b_n) \\  \\

  \dot b _n& =& 2 \sigma _n a_n^2 -2 \sigma _{n-1}  a_{n-1}^2 + (2n+2)  a_n^2 b_n +(2n+1) a_n^2 b_{n+1} + \\
    & &          + (3-2n) a_{n-1}^2 b_{n-1} + (4-2n) a_{n-1}^2 b_n +b_n^3 \ ,
  \end{array}
  \end{equation}

  \smallskip
  \noindent
  with
  \begin{equation}
   \sigma _n = \sum_{i=1}^{n-1} b_i
   \end{equation}
  \smallskip
  \noindent
  and $\sigma_1 =0$.

  \smallskip

  For $n\ge 3$ we define $X_n$ by
  \begin{equation}
  [X_1, X_{n-1} ] = (n-2) X_n \ .
  \end{equation}
  \smallskip
  \noindent
  We consider $X_n$ as an equivalence class of vector fields. We define
  $X_n \sim Y_n$ if $X_n -Y_n = k \chi _{n+1} $, for some real number $k$. It
  can be easily shown that $[X_i, X_j] $ is  equivalent  to
  $(j-i) X_{i+j}$ for $i,j \ge 0$.
  Moreover, we believe, but we don't have a proof, that the two expressions are actually  equal (not just equivalent).

 \bigskip
  \smallskip
  \noindent
  {\bf 3. }  In this section we find an  infinite sequence of evolution vector
   fields that are symmetries of equations (3). We do not know if every symmetry
   of Toda equations is included in this sequence.

   \smallskip

  We  begin by writing  equations (3) in the form

\begin{equation}
\begin{array}{rcl}
 \Gamma _j& =& \dot a_j -a_j b_{j+1} +a_j b_j =0 \\
\Delta _j& =& \dot b_j -2 a_j^2 +2 a_{j-1}^2  =0  \ .
\end{array}
\end{equation}

\smallskip
\noindent
We look for symmetries of  Toda equations. i.e.  vector fields of the
form
\begin{equation}
{\bf v} = \tau {\partial \over \partial t} + \sum_{j=1}^{N-1} \phi_j
{\partial   \over \partial a_j} + \sum_{j=1}^N \psi_j {\partial \over \partial b_j } \
\end{equation}

\smallskip
\noindent
that  generate the symmetry group of the Toda System.
\smallskip
\noindent
The first prolongation of ${\bf v}$ is
\begin{equation}
 {\rm pr}^{(1)} {\bf v} = {\bf v} + \sum_{j=1}^{N-1} f_j { \partial \over
\partial \dot a_j } + \sum_{j=1}^N g_j {\partial \over \partial \dot b_j} \ ,
\end{equation}

\smallskip
\noindent
where
\begin{equation}
\begin{array}{rcl}
f_j&=& \dot{\phi}_j -\dot {\tau} \dot a_j  \\
g_j& = &\dot{\psi}_j -\dot {\tau} \dot b_j  \ .
\end{array}
\end{equation}
\smallskip
\noindent
 The infinitesimal condition for a group to be a symmetry of the system is
 \begin{equation}
 \begin{array}{rcl}
{\rm pr}^{(1)} (\Gamma_j)&   =& 0 \\
 {\rm pr}^{(1)} (\Delta_j)& =&0  \ .
 \end{array}
 \end{equation}

\smallskip
\noindent
Therefore we obtain the equations
\begin{equation}
\begin{array}{rl}
 \dot{\phi}_j& - \dot{\tau} a_j (b_{j+1} -b_j) + \phi_j (b_j-b_{j+1} ) +
a_j \psi_j -a_j \psi_{j+1} =0   \\

 \dot {\psi}_j &- 2 \dot {\tau} (a_j^2 -a_{j-1}^2) -4 a_j \phi_j + 4 a_{j-1} \phi_{j-1} =0 \ .
 \end{array}
 \end{equation}

\smallskip
\noindent
We first give some obvious solutions :

\smallskip
\noindent
{\it i)} \ \ $\tau  =0$,\  $ \phi_j =0 $,\  $\psi _j =1 $.
\bigskip
This is the vector field $X_{-1}$.

\smallskip
\noindent
{\it ii)} \ \ $\tau =-1$, \ $\phi_j=0$, \ $ \psi_j=0$.
\bigskip
The resulting vector field  is the time translation
$-{\partial \over \partial t} $ whose evolutionary representative is
\begin{equation}
 \sum_{j=1}^{N-1} \dot a_j {\partial \over \partial a_j} + \sum_{j=1}^N
\dot b_j {\partial \over \partial b_j} \ .
\end{equation}

\smallskip
\noindent
This is the Hamiltonian vector field $\chi_{H_2} $. It generates a Hamiltonian
symmetry group.

\smallskip
\noindent
{\it iii)} \ \ $\tau = -1$, \ $ \phi_j=a_j$, \ $ \psi_j= b_j$.
\bigskip
Then

\begin{equation}
{\bf v} = -{\partial \over \partial t} + \sum_{j=1}^{N-1} a_j {\partial \over \partial  a_j}
+ \sum_{j=1}^N b_j {\partial \over \partial b_j} =- {\partial \over \partial t} + X_0 \ .
\end{equation}

\smallskip
\noindent
This vector field generates the same symmetry as the evolutionary vector field
\begin{equation}
X_0 + t \chi_{H_2}  \ .
\end{equation}

\smallskip
 We next look for some non obvious solutions. The vector field $X_1$ is not
 a symmetry, so we add a term which depends on time. We try
 \begin{equation}
 \begin{array}{rcl}
   \phi_j& = &-j a_j b_j +(j+2) a_j b_{j+1} +\\
   &            &     t (a_j a_{j+1}^2 +a_j b_{j+1}^2-a_{j-1}^2 a_j -a_j b_j^2 ) \\
   &     &            \\
  \psi _j& =& (2j+3) a_j^2 +(1-2j) a_{j-1}^2 +b_j^2 +\\
      &       & t (2 a_j^2 b_{j+1} +2 a_j^2 -2a_{j-1}^2 a_j -2a_{j-1}^2 b_j ) \
 \end{array}
  \end{equation}

  \smallskip
  \noindent
  and $\tau =0$.

 \smallskip
 \noindent
 A tedious but straightforward calculation shows that $\phi_j$, $\psi_j$
 satisfy (23). It is also straightforward to check that the vector field
 $\sum \phi_j {\partial \over \partial a_j} +\sum \psi_j {\partial \over \partial b_j} $
  is precisely  equal to $X_1 + t \chi_{H_3}$. The pattern suggests that
  $X_n + t \chi_{H_{n+2} } $ is a symmetry of Toda equations. In
  the course of the proof we use some properties of the first three Poisson
  brackets $w_1$, $w_2$, $w_3$  of the Toda lattice.  These three brackets
  have been known for some time \cite{adler}, \cite{kup}.

 \bigskip
  \smallskip
  \noindent
  {\bf Theorem } {\it The vector fields $X_n +t \chi_{n+2} $ are symmetries of
  Toda equations for $n\ge -1$. }

  \bigskip
  \noindent
  {\bf Proof :}  Note that $\chi_{H_1} =0$ because $H_1$ is a Casimir for the
    Lie-Poisson $w_1$ bracket.  We first prove the formula
  \begin{equation}
 [X_1, \chi_l] =(l-1) \chi_{l+1} \ .
 \end{equation}

  \smallskip
  \noindent
  We write $\chi_l =[w_1, H_l]$ where $[\ ,\ ]$ denotes the Schouten bracket.
  We use the super Jacobi identity for the Schouten bracket.
\begin{equation}
[[w_1, H_l], X_1]+[[H_l, X_1], w_1] + [[X_1,w_1], H_l] =0\ .
\end{equation}
\bigskip
\noindent
Therefore,
\begin{equation}
\begin{array}{rcl}
[X_1, \chi_l]& = &(l+1) [H_{l+1}, w_1]  -2[w_2, H_l]   \\

             &= &  (l+1) \chi_{l+1} -2 \chi_l^2 \ .
\end{array}
\end{equation}

\smallskip
\noindent But $\chi_l^2=\chi_{l+1}^1 =\chi_{l+1} $. This is a
Lenard type relation which is easily checked. See \cite{damianou1}
for details. Therefore, we have
\begin{equation}
[X_1, \chi_l] =(l-1) \chi_{l+1}  \ .
\end{equation}

\smallskip
In the same fashion one can  prove that
\begin{equation}
[X_2, \chi_l] =(l-1) \chi_{l+2}  \ .
\end{equation}

\bigskip
 To prove it, one  uses the relation $\chi_l^3 =\chi_{l+1}^2 =\chi_{l+2} $. For
 $n\ge 3$ we use induction on $n$.
 \begin{equation}
 \begin{array}{rcl}
    [X_{n+1}, \chi_l]&=&[ { 1 \over n-1} [X_1, X_n], \chi_l]  \\
                     & =&-{ 1\over n-1} \{ [[X_n, \chi_l], X_1] - [[\chi_l, X_1],X_n]\} \\
                     &=&{1 \over n-1} \{ [X_1, \chi_{n+l}]+ (l-1) [\chi_{l+1}, X_n ] \}  \\
                     & =&{ l-1 \over n-1} [(l+n-1) \chi_{l+n+1} -l \chi_{l+n+1} ] \\
                     &=& (l-1) \chi_{l+n+1} \ .
\end{array}
\end{equation}

\smallskip
\noindent
In particular, for $l=2$, we have $[X_n, \chi_2] =\chi_{n+2} $.

\bigskip
Since the Toda flow is Hamiltonian, generated by $\chi_2$, to
  show that $Y_n =X_n +t \chi_{n+2} $ are symmetries of Toda equations we must
verify the equation
\begin{equation}
{\partial Y_n \over \partial t} + [\chi_2, Y_n] = 0 \ .
\end{equation}

\smallskip
\noindent
But
\begin{equation}
\begin{array}{rcl}
 {\partial Y_n \over \partial t} + [\chi_2,\  Y_n]& = &  {\partial Y_n \over
 \partial t} +[ \chi_2, \  X_n +t \chi_{n+2}   ]  \\
                                  & = &   \chi_{n+2} -[\chi_n,\  \chi_2] \\
        &=& \chi_{n+2} -\chi_{n+2} =0  \ . \\
        & & \ \ \ \ \ \ \ \ \ \ \ \ \ \ \ \ \ \  \ \ \ \ \ \Box
\end{array}
\end{equation}

\vfill
\eject

\end {document}